\documentclass[review]{elsarticle}

\usepackage{lineno}
\usepackage{hyperref}

\journal{Journal of \LaTeX\ Templates}

\begin{document}

\begin{frontmatter}

\title{Using efficient parallelization in Graphic Processing Units to parameterize stochastic fire propagation models}


\author[address1,address2]{M\'onica Denham\fnref{myfootnote}\corref{contacto}}
\ead{mdenham@unrn.edu.ar} 
\author[address1,address3]{Karina Laneri\fnref{myfootnote}\corref{contacto}}
\ead{karinalaneri@gmail.com}

\cortext[contacto]{Corresponding authors}

\fntext[myfootnote]{These authors contributed equally to this work}

\address[address1]{Consejo Nacional de Investigaciones Cient\'icas y T\'ecnicas (CONICET). Argentina}
\address[address2]{Laboratorio de Procesamiento de Se\~nales Aplicadas y Computaci\'on de Alto Rendimiento. Sede Andina, Universidad Nacional de R\'io Negro, Argentina. }
\address[address3]{F\'isica Estad\'istica e Interdisciplinaria, Centro At\'omico Bariloche, R\'io Negro, Argentina. }

\begin{abstract}
Wildfires are a major concern in Argentinian northwestern Patagonia and in many ecosystems and human societies around the world. We developed an efficient cellular automata model in Graphic Processing Units (GPUs) to simulate fire propagation. The graphical advantages of GPUs were exploited by overlapping wind direction, as well as vegetation, slope, and aspect maps, taking into account relevant landscape characteristics for fire propagation. Stochastic propagation was performed with a probability model that depends on aspect, slope, wind direction and vegetation type. Implementing a genetic algorithm search strategy we show, using simulated fires, that we recover the five parameter values that characterize fire propagation. The efficiency of the fire simulation procedure allowed us to also estimate the fire ignition point when it is unknown as well as its associated uncertainty, making this approach suitable for the analysis of fire spread based on maps of burnt areas without knowing the point of origin of the fires or how they spread.
\end{abstract}


\end{frontmatter}

\section{Software availability}
\noindent
Name of software: CUDAFires\\
Developers: Karina Laneri, M\'onica Denham\\
Contact address: John O'Connor 181, San Carlos de Bariloche, R\'io Negro, Argentina.\\ 
Email: karinalaneri@gmail.com, mdenham@unrn.edu.ar \\
Availability and Online Documentation: All the source code is available with a simple request to the authors. \\
Year first available: 2016 \\
Hardware required: CUDA devices with compute capability 1.1 and above. \\
Software required: CUDA SDK \\
Programming language: C/CUDA \\
Program size: 3.0MB. \\

\section{Introduction}
\label{Introduction}
Wildfire affects many ecosystems and human societies around the world. Global environmental change is apparently producing changes in fire regimes, underscoring the importance of understanding how fire spreads in different ecosystems \cite{Stephens2013}. Topography, wind velocity and direction, fuel properties, air and fuel moisture, fire history and topographic aspect are some of the variables that can influence how fire will propagate at a specific site \cite{Abdalhaq_2004,Bianchini_2010,Denham_2012,Denham_2009}. In the Argentine Patagonia, forest and shrubs have high fuel loads (100 tons/hectare or more of rotting wood, fallen trees, etc.). The relative humidity of this coarse material achieves the critical level of 20\% during dry seasons and weather conditions maximizes the fire risks during summer months.  

Several models and simulators have been developed, i.e. BEHAVE, FARSITE, fireLib, PROMETHEUS \cite{Finney1998, Bevins1996} which  implement related models of fire spread based on the physics of fire behavior \cite{Rothermel1972, VanWagner1977, VanWagner1987}. However these forest fire simulators use environmental variables as inputs which are usually difficult to quantify for many specific sites like Patagonia. Therefore the advantage of probabilistic models in our region relies in the fact that fuel models are still not available for the Patagonic region.  For instance, wind speed and direction are hard to measure in a fire context because fire itself produce strong wind gusts. Fuel and air humidity change because of fire. Fuel type is usually considered as the average of typical vegetation of the area, while weather and topography are generally interpolated values that can be calculated for example using WindNinja \cite{Abdalhaq_2004,Denham_2009,SanJuan_2016,Denham_2012}. 

The cellular automaton (CA) approach was efficiently used by Russo et al. \cite{Russo2014} to run thousands of simulations to build hazard maps that have been shown to be robust and efficient in predicting fire spreading behavior in several cases. 
Some stochastic spatial propagation models where fire contagion occurs from cell to cell in a lattice were successfully used to reproduce fire-scar data of semi-arid mountain systems in USA as well as of Northern Patagonia Andean region \cite{Morales2015,Kennedy2010}. In these models the propagation probability from cell to cell is calculated from several covariates that represent the environment in which propagation takes place \cite{Morales2015}, instead of using a function based on the physical principles of heat propagation \cite{Rothermel1972}. The associated parameters of the model can vary in a range of acceptable values. One possible method for the exploration of parameter space is the Genetic Algorithm (GA) which has proven to be useful for Individual Based Model (IBM) calibration  \cite{Vaart2015, Koza92, Abdalhaq_2004,Denham_2009} as well as for other dynamical models \cite{Alonso2010}. 

As a benchmark, the model used in \cite{Morales2015} was run with SELES\cite{Far11} and the performance of one hundred thousand simulations of 9 fires with known ignition point fitted using Approximated Bayesian Computation (ABC), took about 10 days on a PC, running on Intel(R) i7-4770K CPU. This method requires a large number of simulations to explore the parameter space, for which high performance computing can be specially useful. Often, millions of simulations need to be compared with a data set of reference, using a given measure of similarity between simulated and observed data (or fitness in the GA sense). This stage of the process can became a bottle neck of the application.

In this work we develop a high performance application, parallelized using General Propose Graphical Processing Units (GPGPU) architectures and CUDA programming model explained in the \ref{GPGPU-CUDA}, for the type of fire propagation models presented in \cite{Morales2015}. We propose a Genetic Algorithm (GA) in order to improve the search on parameter space. This methodology was previously tested \cite{Denham_2009,Abdalhaq_2004,Denham_2012} for forest fire parameter estimation and was proved to be very effective to accelerate convergence to the best set of parameters. Alternatively, a more simple brute force method based on Monte Carlo sampling (MC) can be used to find the best set of parameters. Albeit computationally more simple, this method is less efficient in terms of computational time, as we will show later.

Other fire simulators, including BEHAVE, FireLib, and FARSITE do not use GPU's and due to their sequential nature, as well as the high amount of input data they need, they are unsuitable for real-time applications for simulation and prediction \cite{Smith2016_Thesis} unless they implement parallel strategies like MPI and OpenMP runtime environments \cite{Cencerrado2014}. Recent papers \cite{Smith2016,Smith2016_Thesis,Sousa2012} show that using the GPUs, accelerations ranging from 64x to 229x were reached and therefore a real-time simulator can become possible, which would be a vast improvement on the current state of the art. However these works do not implement the fitting methodology that we develop in this work. On top of the runtime improvement, interactive simulations on GPUs are actually used as operational prediction tools for fire propagation \cite{Miller2015}.   

The work we present here has two main goals: on the one hand the development of a high performance modular open-source application to accelerate fire propagation prediction and scale-up fire propagation applications. On the other hand, the accurate estimation of fire propagation parameter values from maps of burnt areas. 

An important limitation of the approach presented by Morales et al. \cite{Morales2015} is that the original ignition point has to be known in order to perform the estimation of fire propagation parameters. Very recent works have developed alternative ways to determine the fire ignition point from only the final fire perimeter \cite{Monedero2017}. 
With an efficient simulation procedure like the one presented in this work, we are able to recover all five parameters that characterize our fire propagation using only the information of the final burnt area corresponding to a single fire.
However as expected, there is a limited number of propagation parameters that can be accurately determined from the information of only one fire scar and the treatment of the ignition point as an additional parameter may generate overfitting. Besides this limitations that will be discussed, the feasibility to determine the ignition point and the associated errors, constitutes a valuable information when mapped fires have unknown starting points.

Even though we don't take into account fire evolution in time, our application was built with the potential to fit fuel consumption times as the fire spreads, allowing to test in the future different hypothesis on the inflammability of different vegetation types.

\section{Methods}

\subsection{Forest Fire Spread Simulation}
\label{ForestFireSpreadSimulation}

Fire spreads to neighboring cells according to the following probability (Equation \ref{Eq_fire_probability}) :

\begin{equation}
\label{Eq_fire_probability}
p = \frac{1}{1+\exp(-(\beta_0 + \beta_1 I_f + \beta_2 \psi + \beta_3 \omega + \beta_4 \sigma ))}
\label{Eq:Ppropaga}
\end{equation}

\noindent
where $p$ is the propagation probability that is a function of fuel type (forest or shrubland), aspect, relative wind direction and slope of the target cell. 

In Equation \ref{Eq_fire_probability}, $I_f$ is a vegetation indicator variable equal to 1 for cells occupied by forest and 0 otherwise, so that $\beta_0$ is the baseline fire-propagation probability for shrubland cells and $\beta_1$ measures the difference between shrubland and forest. For instance, if fire is less likely to propagate in forest than in shrublands, then $\beta_1$ should be less than zero. 

Aspect has an important effect in the studied area due to its effects on fuel moisture: sites facing towards the NW have the driest conditions and sites facing SE are the moistest. Thus for landscape cells with a slope greater than 5 degrees, the relative aspect $\psi$ is calculated as: $\psi = \cos(\theta - 315^{\circ}$). Then the relative aspect takes values of 1 for NW facing sites and -1 for those facing SE. 

As in \cite{Morales2015} , wind direction for each cell was derived from the average dominant wind direction during the fire season and modified from topography using WindNinja \cite{Fort09}. 
Then, for each of the 8 cells neighboring an ignited cell $\omega = \cos(\phi_w - \phi_c) $ was calculated where $\phi_w$ is the wind direction in the target cell and $\phi_c$ correspond to the angle between the ignited cell and the neighboring cell. Then, $\omega$ takes values of 1 if the landscape cell is downwind from the ignited cell and -1 if it is upwind from it.  

Slope is calculated based on the digital elevation map (DEM) and is ignored whenever the target landscape cell is at a lower elevation or when slope is zero. The normalized slope ($\sigma=slope/100$) is used when the target cell is higher in elevation than the burning cell. 

Therefore, parameters $\beta_2$, $\beta_3$ and $\beta_4$ modify the propagation probabilities according to aspect ($\psi$), wind direction ($\omega$), and slope ($\sigma$) respectively, being the upper limit for this logistic function equal to 1. The $\beta_i$ values are the same for all cells since they are characteristic of the whole fire propagation process, being the coefficients ($\omega, I_f, \psi, \sigma$) the ones that change from cell to cell according to the environment.   

The model is defined on a two-dimensional lattice of length L that represents the landscape, with sites in four possible states, burnt and active, burnt and inactive, burnable and impossible to burn. A burnt cell gets inactive after a given time that depends on the vegetation type of the cell. For simplicity, we set both times for forest and shrubland equal to one time step of simulation.

To test our method, we generated a synthetic (simulated) fire with a set of known parameters extracted from \cite{Morales2015}, that from now on we'll refer to as the "reference fire".

Fire ignition and propagation were set according to the following rules:
\begin{itemize}
\item For the reference fire, the ignition point was set fixed at an arbitrary site. However, for simulations we considered two cases: fire starting from the same point than the reference fire or fire starting from a random ignition point within the reference fire burnt area. 
\item All cells are checked in parallel and a given target cell is burnt 
according to the probability $1-(1-p_i)^8$ with $i=1,2..8$ the number of neighbors cells and $p$ defined in Equation \ref{Eq:Ppropaga}. 
\item Once the fire propagation stops by itself, we compute the fitness according to Equation \ref{Eq:FitnessDenham} (section \ref{ModelFitting}). We got sure that the fire doesn't reach the edges of the lattice.
\item We ranked all simulations according to their fitnesses. We selected the ensemble of parameters with fitness smaller than a cutoff value. The cutoff was chosen as the value from which the histogram is stationary. 
\end{itemize}

For model calibration we took as the reference a fire produced with a set of parameters extracted from \cite{Morales2015}. We then produced one million simulations with different sets of parameters that were compared to the reference in order to recover the known set of parameters. 

\begin{figure}[h!]
\begin{center} 
\includegraphics[width=1.0\textwidth]{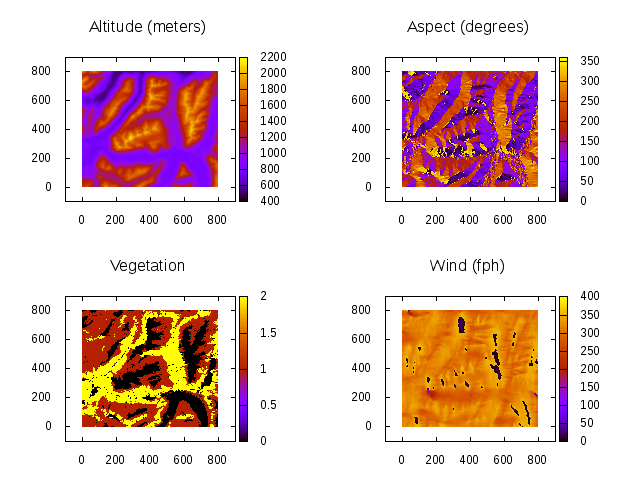}
\caption{Landscape layers where the fire propagation takes place. Wind direction was considered as the average wind in the region. Vegetation values are: 0 for non burnable cell (lakes, roads and firewalls), 1 for forest and 2 for shrubland.}
\label{fig:maps}
\end{center}
\end{figure}

Fire propagation occurs on the substrate shown in Figure \ref{fig:maps} that corresponds to the Northern Patagonia Andean region at $40^{\circ}-41^{\circ}30$'S latitude. Wind direction is highly constant during the fire season with 78\% of day coming from NW or WNW. Terrain is mountainous, with valleys and steep slopes formed by glacial activity. Average annual precipitation varies approximately from 800mm to 3000mm (western limit and eastern areas respectively) and approximately 60\% of the total precipitation falls during the winter season (from May to August). Spring and summer months are typically dry (\cite{Paruelo1998}). 

Raster maps for slope, aspect, vegetation and average wind speed and direction are used as inputs. Maps meta data allow them to be geographically referenced in the physical real space. In our example, maps were 801 x 801 cells, with 30m x 30m cell resolution. Figure \ref{fig:maps} shows some of these maps.

\subsection{Model fitting}
\label{ModelFitting}

A Genetic Algorithm (GA) was used to search for combinations of input parameter values that generate fire scars similar to that of the reference fire. According to the GA methodology, populations are formed by individuals or combinations of $\beta_i$ values (from Equation \ref{Eq_fire_probability}) that are the propagation parameters of the fire simulator. As an alternative more simple methodology we implemented a brute force Monte Carlo algorithm explained in \ref{MC} that allowed us to have a benchmark to compare simulation times between the Genetic Algorithm and a very simple brute force Monte Carlo approach.      

A measure of similarity between simulation maps and reference map is needed to select the best ensemble of simulations that determine parameters and associated uncertainty (see Supplement of \cite{Alonso2010}). 

Two fitness functions were used, both of them based on the difference between simulated and reference fire maps (Equation \ref{Eq:FitnessDenham} and the other one is presented in \cite{Morales2015}).  The "reference fire" of our synthetic experiment was compared with simulations using both fitness functions. Given that our fire propagation is stochastic, the same set of parameters and ignition point will give rise to different fire-scars, for different random seeds. 

One of the fitness functions (Equation \ref{Eq:FitnessDenham}) implemented and used in \cite{Denham_2009} denotes the discrepancy between real and simulated fires, counting burnt overlapped cells. This function is 0 when simulation and real maps are perfectly overlapped and does not discriminate cells by fuel type. 
 
\begin{equation}
\label{Eq:FitnessDenham}
\delta = \frac{\left(A \cup A^* \right) - \left(A \cap A^* \right)}{A}
\end{equation}

\noindent
In Equation \ref{Eq:FitnessDenham} A denotes reference map (real fire map) and $A^*$
denotes simulated fire map. The $\cup$ symbol means the amount of cells in the union of burnt cells (burnt cell in simulated map or burnt cell in the reference map), $\cap$ is the total of burnt cells in both maps (the intersection of the reference and simulated map). 
The numerator of Equation \ref{Eq:FitnessDenham} is the difference between real and simulated map. The denominator is for normalizing that difference with respect to the size of real fire area and accounts for the amount of burnt cells in the real fire map.\\ 
The other fitness equation was previously defined elsewhere \cite{Morales2015}. Both fitness functions were used. Simulations were ranked according to the fitness and the corresponding parameter values were used to estimate parameters and associated uncertainties. With the best ranked simulations we plotted the histograms of $\beta$ values that allowed to determine the uncertainty associated with the fitting methodology.
Simulations were parallelized in order to take advantage of actual hardware architectures like graphic cards. Next section presents the CUDA C parallel implementation of our application. 

\subsection{CUDA C Implementation}
\label{CUDA_Implementation}

Our parallel application was developed in CUDA C for General Purpose Graphical Processing Units (GPGPUs). These graphic cards are a massively parallel platform that offer high computational power. \ref{GPGPU-CUDA} describes the main features of graphic cards and CUDA parallel model. 

Our approach is based on the parallelization of simulation steps: a thread per matrix cell is launched in order to compute if a (target) cell will be ignited or not. Each thread calculates Equation \ref{Eq_fire_probability} for each of its neighbouring cells and accumulates the product $(1-p_i)^8$ which is the probability of no ignition of the target cell. Therefore the ignition probability of the target cell is calculated as $1-(1-p_i)^8$. A random number in the range [0,1] is calculated and compared with the ignition probability determining if fire spreads to the target cell. Therefore, the state of all cells are updated at the same time, avoiding sequential cell by cell computation and in this way, reducing total runtime. 

Furthermore, in order to reduce runtimes, a GA improvement was implemented: for each simulated map, when fire stops the current simulation is terminated and next simulation starts. This is done with a flag used in GPU but tested in CPU, since the cellular automata is driven and executed on GPU. This data movement from GPU to CPU, produces an overhead that was not significant in this case because most simulations stopped at early iterations. Then, this reduction of simulation steps overcomes the communication overhead.  

As we are dealing with stochastic simulations, even if the set of parameters for all the simulations is exactly the same as the reference parameters, we obtain different final scars for different random seeds. The fitness difference for that case is around 0.05, which we took as a baseline fitness below which we have the effect of stochasticity.
For histograms of fitted propagation parameters with fixed ignition point, we took the cutoff value of 0.3, while for the case of variable ignition point the cutoff value used to plot the histograms was 1.5. The cutoff values were chosen as the value from which the histogram shape was stationary.

We had used a counter based random number generator: the Random123 library. This library is a collection of counter-based random number generators (CBRNGs) for CPUs (C and C++) and GPUs (CUDA and OpenCL).  From this library, Philox generator was used. It allows to generate a very large number of random numbers concurrently, and guarantees that the random numbers are uncorrelated and uniform distributed between 0 and 1.

\section{Results and discussion}
\label{Results}
Parameters of the fire of reference are listed in Table \ref{Tab:FitPosterior} and were chosen so that fire scars never reach the borders of the lattice. Propagation stops because of the environment and because cells have a turn off time that depends on vegetation fuel type. We present the results for a turn off time of 1 step of simulation for both forest and shrubland. Parameters corresponding to forest, wind, slope and aspect were randomly sampled from a uniform initial distribution and from a gaussian distribution. In order to provide clarity through this section, only results using an initial gaussian distribution are shown and results using an initial uniform distribution can be seen in \ref{Sec:moreResults}. 

\begin{table}[h!]
\begin{center}
\caption{Initial distribution features of all $\beta_i$ input parameters of the model.}
\label{Tab:Prior_characteristics}
\begin{tabular}{cccc}
 \textbf{Initial} & \textbf{Mean} & \textbf{Standard} & \textbf{Range} \\
\textbf{Distribution} &  & \textbf{Deviation} & \\
\hline
\noalign{\smallskip}
Gauss & 0 	& 5 & $(-\infty,\infty)$	\\
Uniform & 0 & 17.3 & $[-30,30]$	 	\\
\end{tabular}
\end{center}
\end{table}

We show in Table \ref{Tab:Prior_characteristics} the initial distribution parameters. 

\begin{figure*}[h!]
\begin{center}
\includegraphics[width=0.9\textwidth]{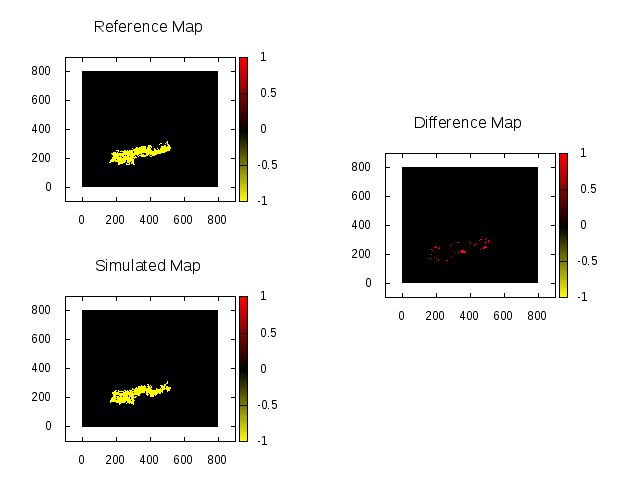}
\caption{Difference (right panel) between reference fire (upper left panel) and the best simulation (lower left) with all the fires starting from the same ignition point (x=400; y=250). Fitness was 0.15 (calculated with Equation \ref{Eq:FitnessDenham}).}
\label{Fig:DifIgniFix}
\end{center}
\end{figure*}

As an example, Figure \ref{Fig:DifIgniFix} shows the difference map between the best of all simulations and the reference map after fitting and estimating the best parameter values. Both simulations started from the same ignition point. 

\begin{figure*}[h!]
\begin{center}
\includegraphics[width=0.9\textwidth]{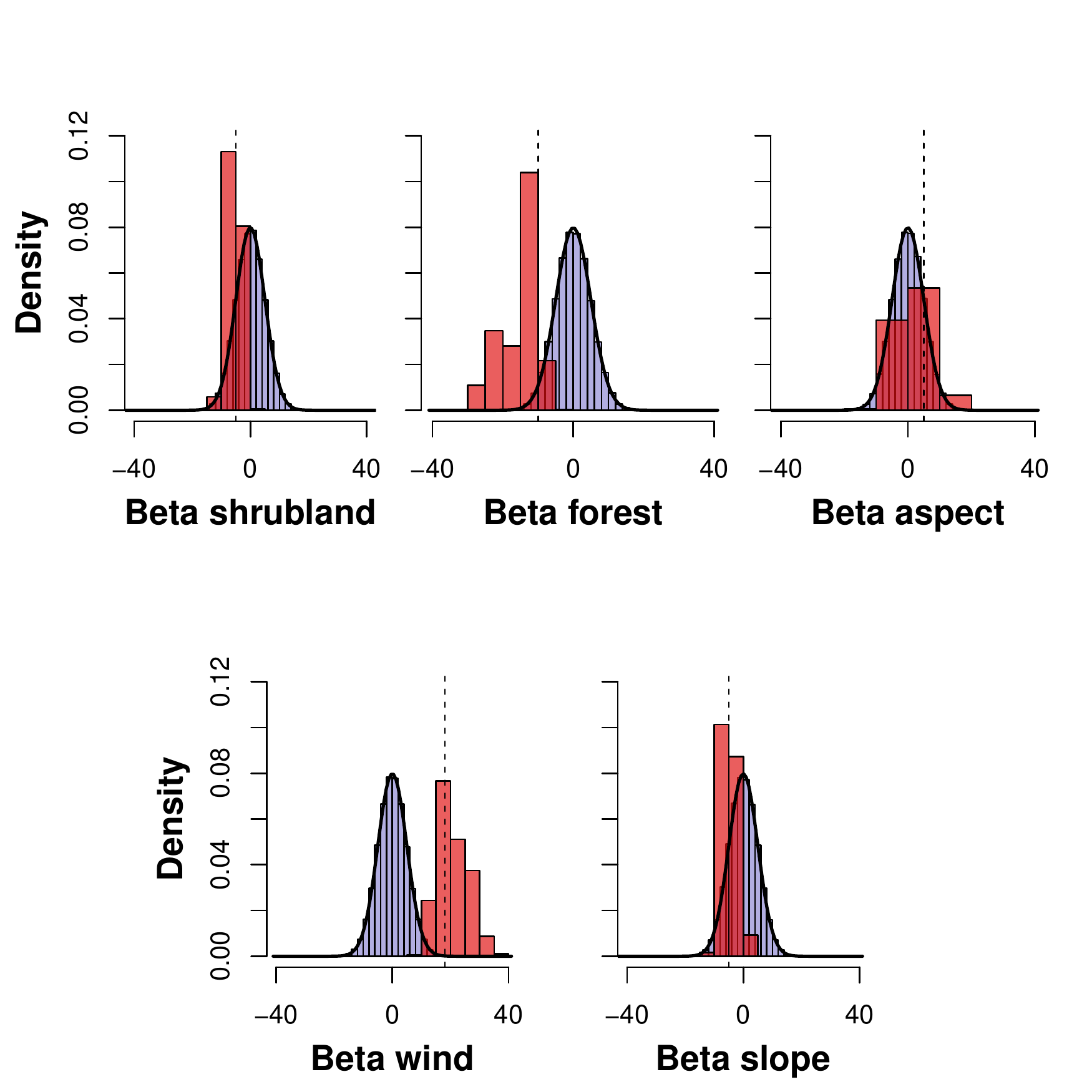}
\caption{Histograms for $\beta_i$ values after fitting the model using the Genetic Algorithm (red bars). Dotted line indicates the true value of the corresponding parameter that was used to generate the fire of reference. One million simulations were performed with all fires starting from the same ignition point (x=400; y=250) and with propagation parameters taken from an initial Gaussian distribution (mean=0, sd=5; solid black line and light blue bars). Fitness was calculated with Equation \ref{Eq:FitnessDenham}.}
\label{Fig:GaussFMIGNIFIX}
\end{center}
\end{figure*}

After performing one million simulations starting from the same (known) ignition point, sampling parameter values from an initial gaussian distribution, we obtained histograms of values for each of the parameters using genetic algorithm (Figure \ref{Fig:GaussFMIGNIFIX} and Table \ref{Tab:GaussFMIgniFix}). Similar results are obtained when sampling from a uniform initial distribution (Figure \ref{Fig:IgniFixFM} in  \ref{Sec:moreResults}) showing the independence of the methodology on the initial distribution. 

It's a remarkable result to recover the all five parameters of fire propagation, by only comparing between the final fire scar of reference and simulations. For instance, we are dealing with a 5 dimensional search space that implies a very large amount of simulations. The Genetic Algorithm is a search strategy that might not work for some situations, e. g. the search may get stuck in a relative minimum of the likelihood surface or it might not work if the surface is very complex with lots of minimums. Also if there are interactions between parameters a distribution with more than one maximum can be obtained, which indicates that more than one set of parameters could give place to the same observed fire-scar.

\begin{table}[h!]
\caption{Parameters of the reference fire and after fit histogram values. The $\beta$ parameters are defined in Equation \ref{Eq:Ppropaga}. The ignition point coordinates were fixed at the values of the reference fire. The average parameter values were calculated from 62732 (initial gaussian) and 70516 (initial normal) simulations that were below the fitness cutoff value of 0.3.}
\label{Tab:GaussFMIgniFix}
\begin{tabular}{cccc}
& \textbf{Reference} & \textbf{Initial Uniform} & \textbf{Initial Normal} \\
\textbf{Parameter} & \textbf{value} & \textbf{mean(sd)} & \textbf{mean(sd)} \\
\hline
\noalign{\smallskip}
$\beta_0$ & -5 	& -8.7(2) & -4.9(2) \\
$\beta_1$ & -10 & -22(6)  & -16.2(6) \\
$\beta_2$ & 5 & -5.1(7)  & 0.4(7) \\
$\beta_3$ & 18.1 & 27(2) & 21.3(6) 	\\
$\beta_4$ & -5 & -9.2(10) & -5.3(3) \\
\end{tabular}
\end{table}

\begin{figure*}[h!]
\begin{center}
\includegraphics[width=0.9\textwidth]{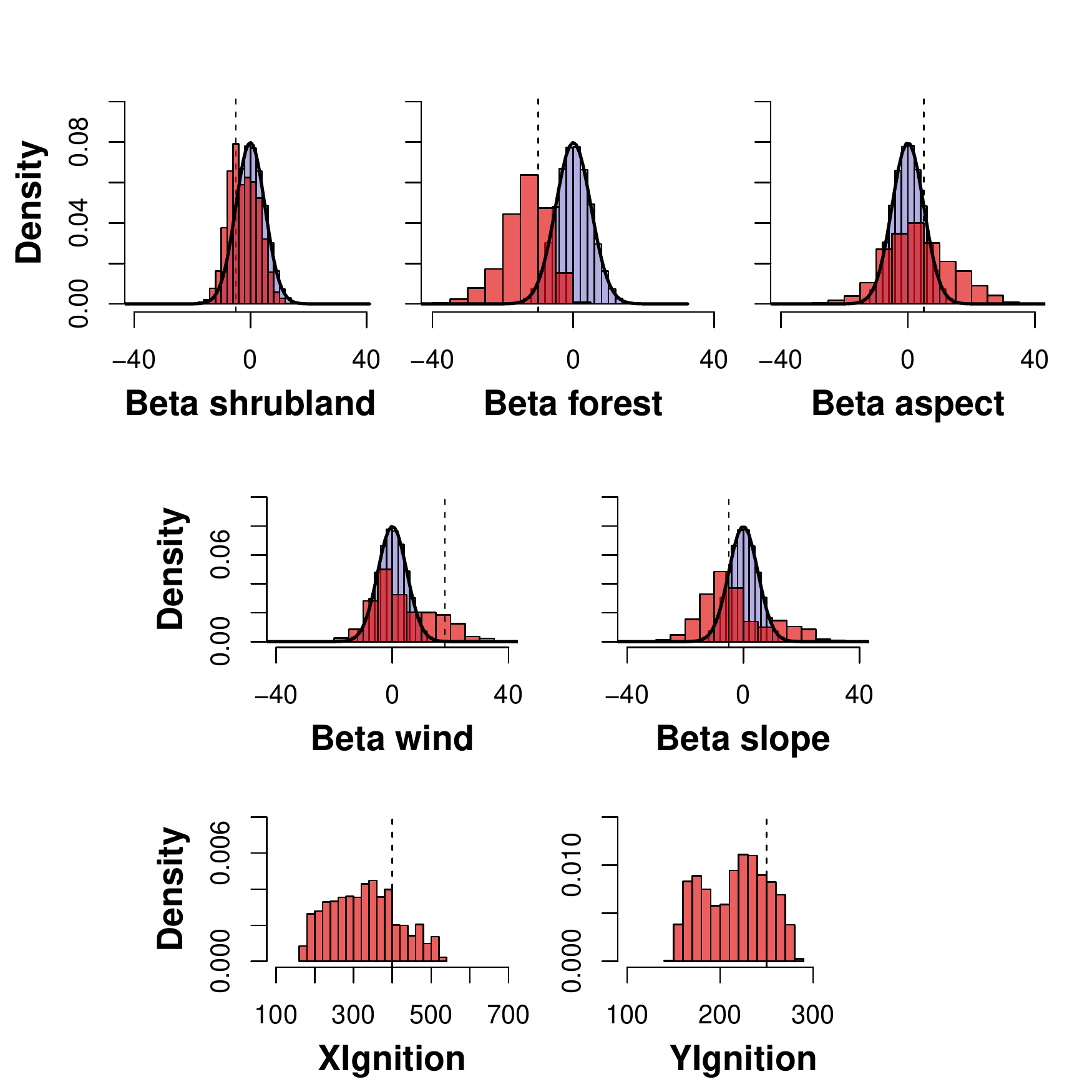}
\caption{Histograms for $\beta_i$ values and ignition point after fitting the model using the Genetic Algorithm (red bars). Dotted line indicate the true value of the corresponding parameter that was used to generate the fire of reference. One million simulations were performed with parameters taken from an initial Gaussian distribution (mean=0, sd=5; solid black line, light blue bars). Fitness was calculated with Equation \ref{Eq:FitnessDenham}.}
\label{Fig:TodovarGaussFM}
\end{center}
\end{figure*}

As already mentioned, it's could be of interest to also determine the ignition point. To that aim we considered the ignition point as an additional parameter, performing one million simulations starting from different ignition points picked at random from the burnt area of the reference fire. As can be seen, histograms after fitting, contain the true parameter values (Table \ref{Tab:FitPosterior}). For the wind parameter ($\beta_3$ in Table \ref{Tab:FitPosterior} ) it's seen that a uniform initial distribution leads to more accurate results (Figure \ref{Fig:IgniVarFM} in  \ref{Sec:moreResults}). However for the rest of parameters a more informative initial distribution seems to be a more appropriate choice (Figure \ref{Fig:TodovarGaussFM}). In other words, the wind parameter (for initial gaussian distribution) or aspect and slope parameters (when an initial uniform is used), loose identifiability when the coordinates of the ignition point are considered as an additional parameter (see Figures \ref{Fig:TodovarGaussFM} and \ref{Fig:IgniVarFM}). 

\begin{table}[h!]
\caption{Parameters of the reference fire and histogram of fitted parameter values using GA. The $\beta$ parameters are defined in Equation \ref{Eq:Ppropaga} and (x,y) is the ignition point coordinates. The average parameter values were calculated from 134253 (initial gaussian distribution) and 141249 (initial normal distribution) simulations that were below the fitness cutoff value of 1.5.}
\label{Tab:FitPosterior}
\begin{tabular}{cccc}
& \textbf{Reference} & \textbf{Initial Uniform} & \textbf{Initial Normal} \\
\textbf{Parameter} & \textbf{value} & \textbf{mean(sd)} & \textbf{mean(sd)} \\
\hline
\noalign{\smallskip}
$\beta_0$ & -5 	& -9(2) & -2.3(5) \\
$\beta_1$ & -10 & -20(6)  & -13.3(6)	 	\\
$\beta_2$ & 5 & -2(19)  & 3.3(10.5)	 	\\
$\beta_3$ & 18.1 &  19(14) & 3.7(10) 	\\
$\beta_4$ & -5 & -4(15) & -2.7(11) 	\\
$x$ & 400 & 331(95) & 328(87)	\\
$y$ & 250 & 224(33)& 217(34)	\\
\end{tabular}
\end{table}

Regarding the efficiency of Genetic Algorithm (GA) as compared with a simple Monte Carlo brute force method (MC), we implemented an alternative and more basic fitting procedure \ref{MC}, that we can take as a benchmark for time comparison. The beta histograms obtained with MC are shown in Figs. \ref{Fig:MC_UnifIngiVar} and \ref{Fig:MC_UnifIngiFix}.
According to our results, the averaged speed of GA is half the one of a simple MC brute force approach, as can be seen in Fig.\ref{Fig:TimesSims_MC_AG}, where we show the histograms of times for each simulation performed with both methods. This is because the ensemble of simulations using GA remains closer to the "real simulation" (specially for the last evolution of the algorithm) while simulations with random sampled parameters are more heterogeneous and therefore have more variability in simulation time.

\subsection{Scalability Analysis}
\label{Sec:scalability_analisys}

When applications with high computation requirements are designed and programmed, is usual to develop them taking into account hardware features, in order to achieve an efficient use of all resources in that architecture (high performance applications). But constant advances in technology requires a software that remains useful and efficient when computer capabilities increase. 

CUDA model and GPU hardware offer high scalable platforms, meaning that CUDA applications may remain efficient through different graphic cards. One of the reasons is that GPU schedulers deliver workload through Streaming Multiprocessors, looking for best occupancy of cores and special units within a multiprocessor. 

We therefore executed our application in different GPUs platforms in order to study its behavior and scalability. Table \ref{Tab:scalability} shows runtimes of our application when different graphic cards are used. In this test the overall lattice size remains constant (map size 801x801 and 1million simulations) as the GPU platform changes.

\begin{table}[h!]
\caption{Additional tests changing hardware architecture. Map size is 801x801 and runtime corresponds to $1x10^6$ simulations, for Genetic Algorithm (GA) and simple brute force Monte Carlo (MC).}
\label{Tab:scalability}
\resizebox{12cm}{!} {
\begin{tabular}{clccccc}
\noalign{\smallskip}\hline
 &  &  \textbf{CUDA} & \textbf{Processing} & \textbf{Memory} & \textbf{Runtime GA}&\textbf{Runtime MC}\\
  & \textbf{Name} & \textbf{Cores} & \textbf{Power*} & \textbf{(GB/s)**} & \textbf{(hs)}& \textbf{(hs)}\\
\hline
\noalign{\smallskip}
1 & Tesla C2070 (2.0) 	&  	 	448 	& 1030  & 144 & 23.39 & 78.74\\
2 & Tesla K40c (3.5) & 	2880	& 4291 $-$ 5040 & 288 & 18.29& 61.60\\
3 & GeForce GTX 780 (3.5)& 	 	2304 	& 3977 & 288 & 15.39& 48.86\\
4 & GeForce GTX Titan (3.5)	&  	2880 	& 5121 & 336 & 12.98& 47.50\\
\noalign{\smallskip}\hline
	\multicolumn{6}{l}{Notes: *Single precision GFLOPs peak. ** Memory bandwidth}\\
\end{tabular}
}
\end{table}

Four different platforms were used in order to evaluate application scalability. Runtimes show us that our application is scalable. When better graphic cards are used, application runtimes are shorter. It is an important feature that no modification is applied over the computer code, it's just compiled and launched through these different architectures. 

We can see that best runtimes were achieved using the GeForce GTX Titan. This graphic card offers the best memory bandwidth (fourth column). When application profiling was used in order to evaluate application performance, we could see that main kernels are memory bounded, that means that memory accesses (for read or write operations) are the limiters of the application. 

Topography, wind, fuel, fire propagation maps are allocated in GPU global memory. This memory is the most expensive memory in the GPU for an application because it has the highest latency, is not cached and the bandwidth is low when not aligned and not coalesced accesses are carried out. 

In order to analyze application performance, two profilers were used: NVIDIA Visual Profiler (nvvp) and NVIDIA Profiler (nvprof) \cite{che14}. Using these profiling tools we could see that read from and write to global memory are very inefficient operations. Using \texttt{gld\_efficiency} and \texttt{gst\_efficiency} metrics we could see that very low percentages of bandwidth use were achieved. This fact means that most of the memory transaction had to be replayed and each access is handle by more than one memory transaction (because access is not coalesced nor aligned). 

For this reason, the use of constant memory and shared memory was analyzed. Both of them are cached memories, then, if read and write operations are studied in order to fit some access pattern, access operation has no latency and are similar to internal register access. But current graphic cards have very small shared and constant memories that are not enough for our application, where several raster maps are necessary to perform simulations. 

\begin{figure*}
\begin{center}
\includegraphics[width=0.6\textwidth]{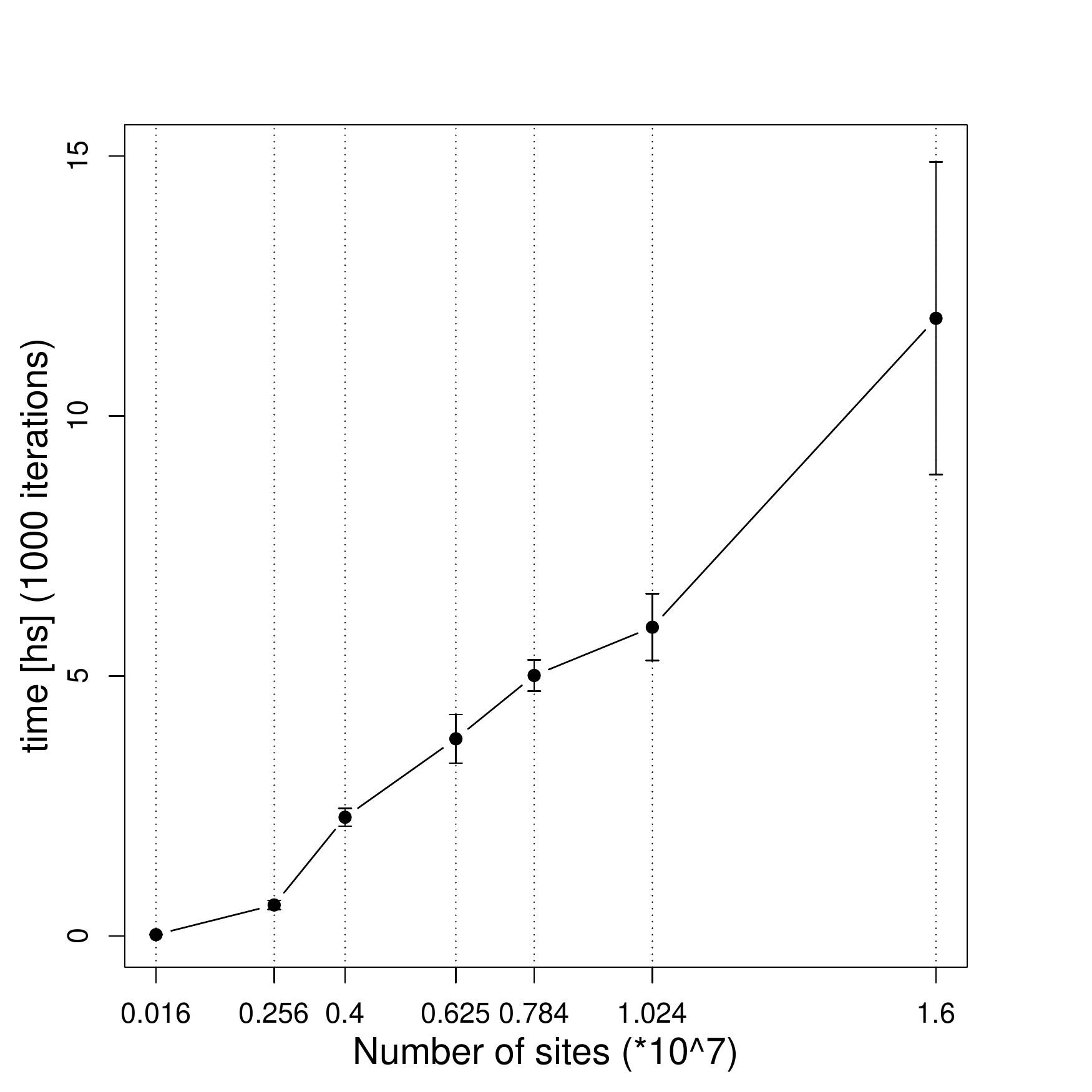}
\caption{Scalability study. Different runtimes for different configuration of fire maps for 10 thousand simulations using the Genetic Algorithm. Vertical lines indicate the total amount of cells (rows x columns).}
\label{Fig:ScalabilityII}
\end{center}
\end{figure*}

Another scalability analysis is the study of the behavior of application runtime when lattice size increases.  
With this objective, more tests were done where fire landscape sizes were consistently varied (fire map, topography map, wind maps, etc.).  Results are showed in Figure \ref{Fig:ScalabilityII}, where runtime (Y axis) is specified for different map sizes (X axis). 

All tests were performed using the Tesla K40c architecture due to memory requirements. This architecture has 12GB of total on board memory while the remaining graphic cards available have 6GB, which was not sufficient to perform all the tests. 

Figure \ref{Fig:ScalabilityII} shows that runtime increases in a linear way when map sizes increases. Even though application maps are squared data structures, and data increases in a quadratic form too, runtimes have a linear growth. This is an important advantage of this parallel solution as compared to the frequently used sequential cellular automata. 
An important result is achieved when 1600x1600 and 3200x3200 cell maps are used for which a reduction in time is observed because maps sizes are multiple of 32. When GPUs are used, warps of 32 threads are executed in a SIMD (Single Instruction Multiple Data) way with better GPU cores occupancy as well as more efficient global memory access.
Therefore it is very important to make a correct design of the application when graphic cards are used in order to obtain high performance applications.
Using the sequential solution developed on SELES as in \cite{Morales2015} it takes near 10 days to perform $9x10^5$ simulations. With our parallel implementation it takes 13 hours to perform 1 million of simulations (GTX Titan Black, GK110 architecture).

\section{Conclusions and open lines}
We have developed an application for forest fire simulation and fitting to accomplish with two main goals: the estimation of parameters involved in fire propagation and the efficient implementation of a parallel modular open source computing tool.

It is of importance to be able to determine stochastic fire propagation parameters and ignition points once a fire had occur. Having that information it's possible to identify which are the more sensitive parameters on the propagation of one or more fires of interest and eventually predict fire propagation under different scenarios.
Our parallel cellular automata is an spatial stochastic model for fire propagation mounted on several layers that describe topography, fuel type, wind speed and direction, vegetation aspect and slope, that can be easily adapted to include other layers, as well as to implement different rules for fire propagation. 

We show in this work that genetic algorithm allows to recover propagation parameter values that generated the reference fire. We proved that departing from an initial shifted distribution we can also recover the known parameters by efficiently surfing in the multidimensional parameter space. We obtained well defined stationary histograms after fitting one million simulations, which makes parallel GPU computing a valuable tool specially when large lattices are used.

The reduction of our application runtime is very significant when compared with previously used serial applications and 
the scalability analysis allows to determine the time needed to run the propagation on any desired lattice size. Depending on the GPGPU technology used, our first parallel version takes between 13 and 23 hours to perform and fit one million of simulations to a given fire of reference using the genetic algorithm.  

Using CUDA profilers we could see that our application is a memory bounded application, then, the study of memory access patterns and optimization opportunities are open lines. In a near future we'll study the application memory use in order to reduce memory latencies as well as to improve memory bandwidth usage. 

With this tool is also possible to estimate the fire starting point, allowing the analysis of mapped fires with unknown starting points. However, some of the propagation parameters loose identifiability when the coordinates of the ignition point are considered as an additional parameter. This could be probably improved either by fitting several fires scars instead of only one (assuming they were generated with the same set of propagation parameters) considering ignition points as nuisance parameters, or by using an ensemble of several fitness functions to rank simulations.

The next step will be to use our methodology to fit real fire scars maps to estimate propagation parameters and eventually fire ignition points, as well as to test other functional dependencies of fire propagation probability. Additionally it would be of interest to fit times of fuel consumption, a feature that will be also added to our application.

Given that this tool is inherently parallel it could be eventually used for fire propagation prediction for which computing times should be the minimum possible simulation times. These results have brought us closer to the real time requirements for forest fire spread prediction which is not our actual main goal but that is a desirable property of any application and should be explored in the near future. 

Taking advantage of graphic cards capabilities we are now developing an interface for the visualization of the fire progress in simulation time to be able to interact with the simulation for management purposes. 
The advantages of GPU's are not restricted to spatial fire propagation but also to a broad range of ecological models that could be studied inspired on a similar program structure, allowing to simulate and fit several models at several scales, specially for large spatial extensions that were difficult to explore with previously used technologies.

\section*{Acknowledgements}
M. Denham and K. Laneri are members of CONICET. M. Denham and K. Laneri are part of the project  TIN2014-53234-C2-1-R of Ministerio de Ciencia e Innovaci\'on (MICINN-Spain). We thank David Alonso and Alejandro Kolton for fruitful discussions.

\section*{Bibliography}

\bibliographystyle{plain}  


%
%
%
%

\appendix
\section*{Appendices}
\section{GPGPU - CUDA model}
\label{GPGPU-CUDA}
Originally GPUs were developed for the video game industry. These cards were designed to achieve high performance in video game applications, where many similar simple computations had to be done in parallel, thus several basic cores were used in an independent graphics chip (deemed Graphics Processing Unit, GPU) in order to alleviate the main CPU load. As their computational power grew up while keeping prices low, GPU became an attractive architecture for High Performance Computing. Simple and dedicated cores became complex and general purpose cores. First generation GPU cores could perform some specific operations suited for the graphical pipeline. Current GPUs cores have grown in complexity and can perform a wide set of operations. This revamped architecture is called GPGPU (General Purpose GPU) \cite{Pic11} \cite{Kir10} \cite{Coo13}. 

The broader use of GPGPU architectures was motivated by the creation and definition of several tools for developing and programming GPGPU applications. Perhaps the most extended of such development environments is CUDA (Compute Unified Device Architecture), which is a language (extensions of other languages such us C, C++, Phyton, etc), a compiler and a programming model \cite{Sand10} \cite{Kir10} \cite{Far11} \cite{CUDAC} \cite{Coo13} \cite{che14}.  

\begin{figure*}
  \includegraphics[width=1\textwidth]{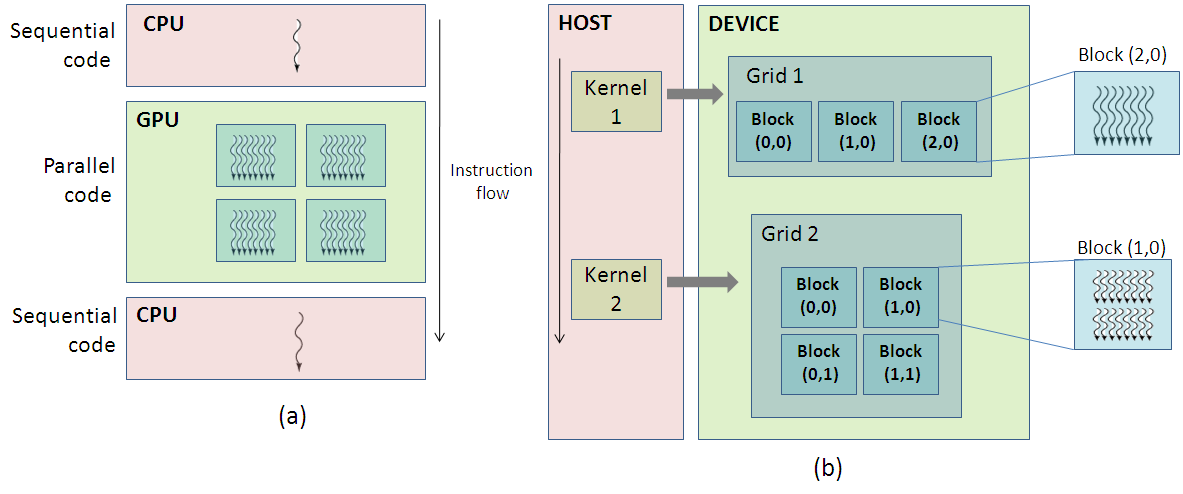}
\caption{(a) Hybrid program CUDA model \cite{CUDAC}. (b) Kernel arrange example \cite{CUDAC}.}
\label{fig:CUDA}
\end{figure*}

Graphic cards are used in conjunction with a CPU, which governs GPU execution. GPU applications are hybrid programs combining sequential and parallel code. Sequential code is executed on CPU and parallel code is executed on the graphic card. Figure \ref{fig:CUDA} (a) shows the execution flow of such a hybrid CUDA program. 

GPU architectures have a memory hierarchy, where each level has its own access velocity, latency, bandwidth, most efficient access pattern, etc. It is necessary to take this hierarchy into account in order to achieve high performance applications. 

Parallel CUDA code is executed in kernels. A kernel is a function that is executed by several threads. These threads are arranged in arrays of 1, 2 or 3 dimensions (figure \ref{fig:CUDA} (b)). Launching a kernel implies the creation of large amount of threads. Thread layout is another important design key for achieving high performance. 

\section{Parallel Genetic Algorithm}
\label{GA}

The search is performed on a large parameter space: combinations of 5 parameters are evaluated.  When the initial ignition point is unknown, the application deals with 7 parameters (X and Y values are added).  

In order to accelerate convergence, a Genetic Algorithm (GA) was implemented. This algorithm is based on three main operators: selection, crossover and mutation. The goal of these operations is to generate populations with individual characteristics with good fitness \cite{Koza92}.
Selection is based on: individuals that are well suited to its environment can survive and inherit its genes to their offspring.
The selection function is based on giving more chances of being chosen to individuals which are well adapted to its environment. Therefore, good features (genes) are inherit by children (individual of next population).
CUDA implementation of selection function includes the elitism function: every evolution includes best individuals of previous citizen. Elitism warranties that good individuals are always selected and its features are included in future generations. Tournament selection was paralellized and implemented using CUDA: a thread for each new individual is launched and each thread chose randomly 2 individuals. Tournament selection involves running tournaments among these two individuals (or "chromosomes"). The winner of each tournament (the one with the best fitness) is selected for crossover. 

\begin{figure*}
\begin{center}
  \includegraphics[width=0.5\textwidth]{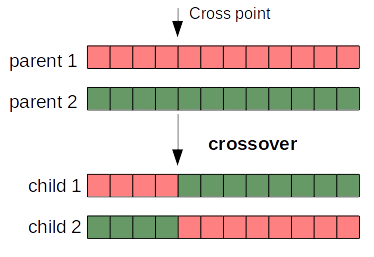}
\caption{One-point crossover of genes.}
\label{fig:crossover}
\end{center}
\end{figure*}

Crossover operator is used in order to mix parents features and to form new individuals. Half of population size threads are launched and each of them choose randomly a cross point. All data beyond that point in either organism string is swapped between the two parent organisms. The resulting organisms are the children. This technique is called one-point crossover, due to just one cross-point is used (Figure \ref{fig:crossover}).

\begin{figure*}
\begin{center}
  \includegraphics[width=0.6\textwidth]{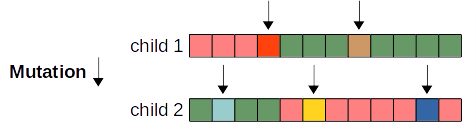}
\caption{Mutation changes makes a tiny  (random) gene change.}
\label{fig:mutation}
\end{center}
\end{figure*}

Mutation operator is used to maintain population diversity. Using a small probability, each of the individual gene can be mutated. This operation avoid that search process falls down in a local maximum or minimum. 
Mutation operation is executed by a thread for each population member. For each individual gene, a random number is compared with the mutation probability. 
We had implemented a dynamic mutation: mutation probability decreases as GA advances. Earlier generation mutation allows to search through more distant zones of the whole search space. Then, as population evolves, it is less likely to have a gene mutation (Figure \ref{fig:mutation}).

\section{Parallel Brute Force Monte Carlo}
\label{MC}
\begin{figure*}[h!]
\begin{center}
\includegraphics[width=0.9\textwidth]{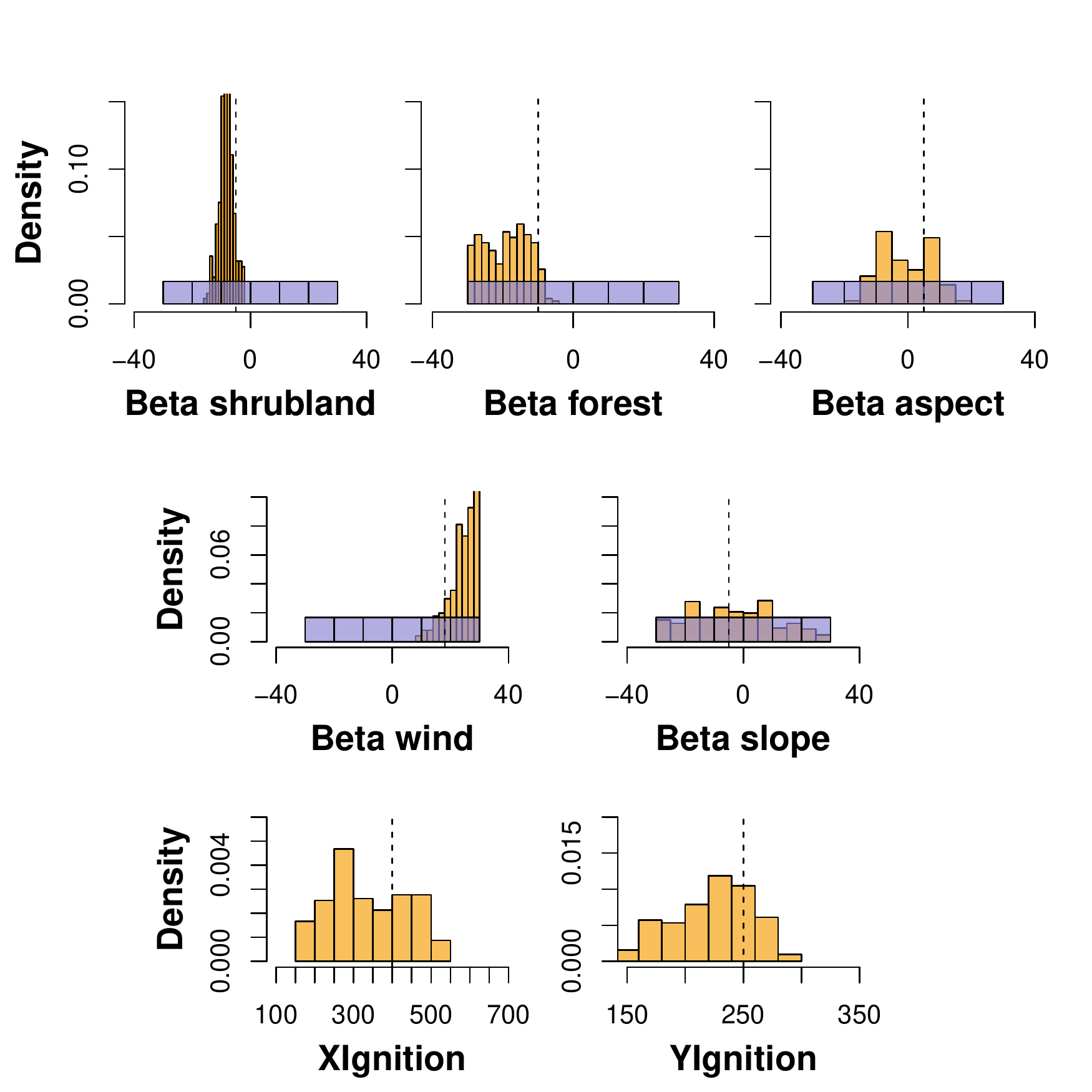}
\caption{Histograms for $\beta_i$ values after fitting the model using brute force Monte Carlo (yellow bars). Dotted line indicate the true value of the corresponding parameter that was used to generate the fire of reference. One million simulations were performed with all fires starting from the same ignition point (x=400; y=250) and with propagation parameters taken from an initial Uniform distribution (mean=0, sd=5; light blue).}
\label{Fig:MC_UnifIngiVar}
\end{center}
\end{figure*}

\begin{figure*}[h!]
\begin{center}
\includegraphics[width=0.9\textwidth]{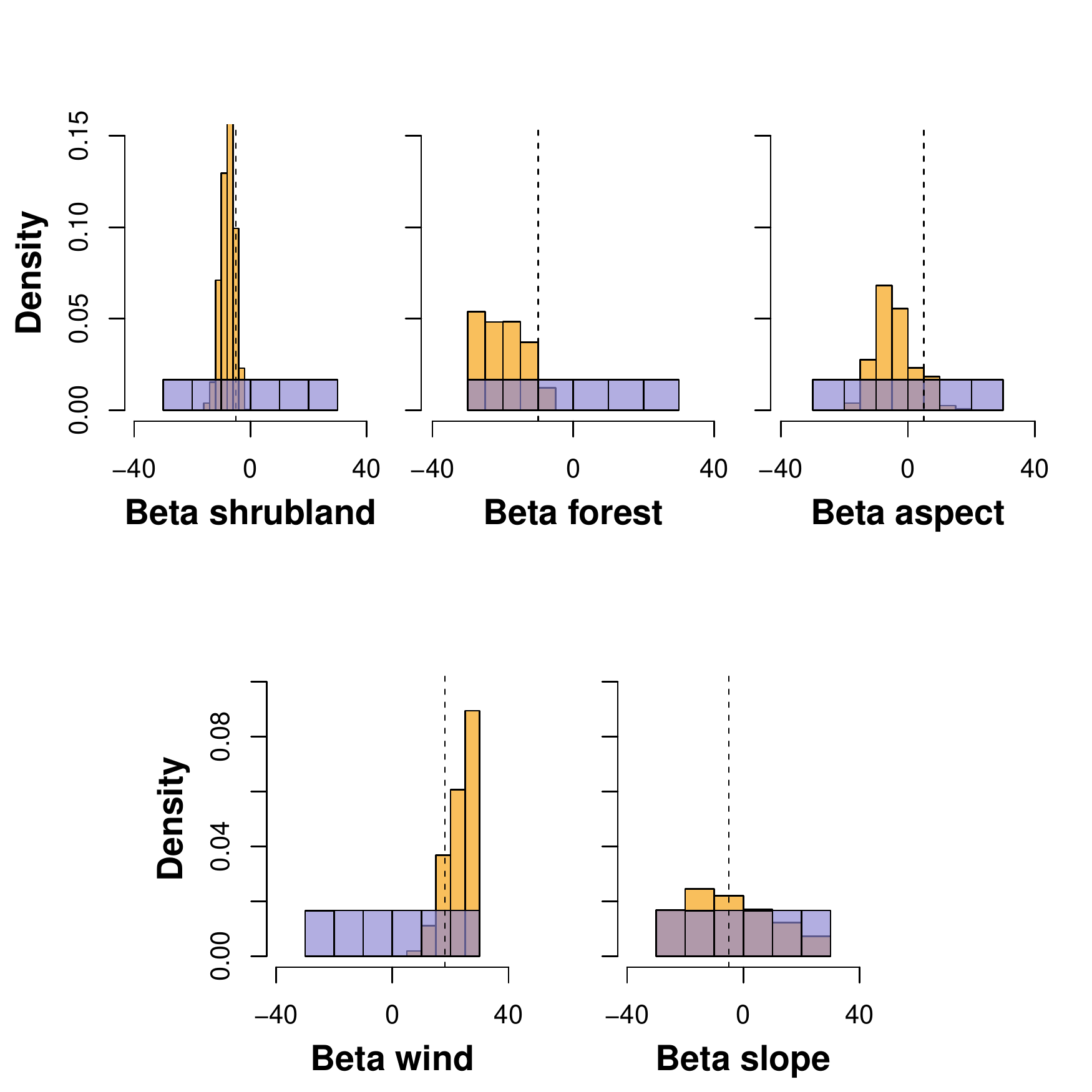}
\caption{Histograms for $\beta_i$ values after fitting the model using Monte Carlo (yellow bars). Dotted line indicate the true value of the corresponding parameter that was used to generate the fire of reference. One million simulations were performed with all fires starting from the same ignition point (x=400; y=250) and with propagation parameters taken from an initial Uniform distribution (mean=0, sd=5 ; light blue).}
\label{Fig:MC_UnifIngiFix}
\end{center}
\end{figure*}

\begin{figure*}[h!]
\begin{center}
\includegraphics[width=0.9\textwidth]{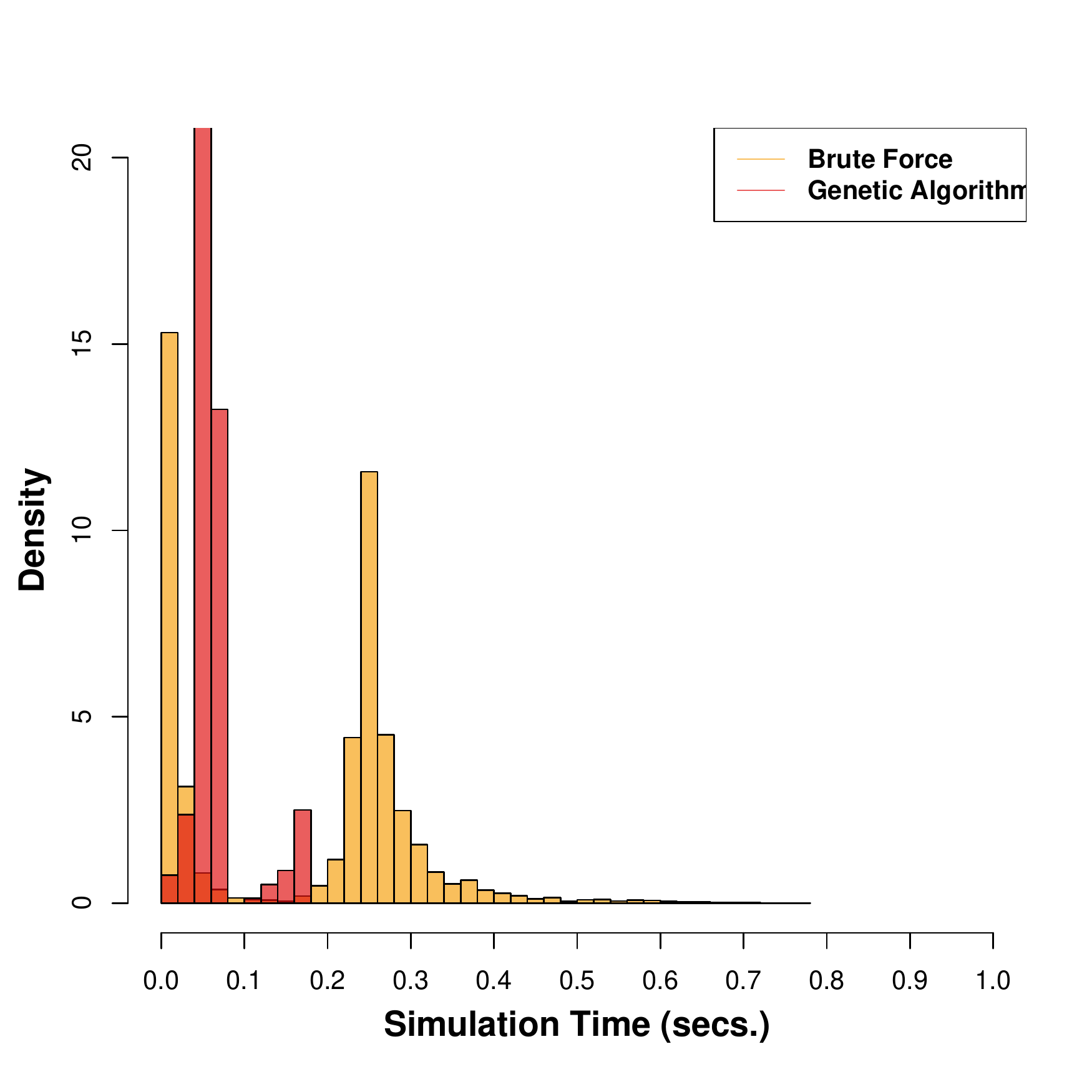}
\caption{Histograms for simulation times using Genetic Algorithm (red) and brute force Monte Carlo (yellow) obtained using a GeForce GXT Titan.}
\label{Fig:TimesSims_MC_AG}
\end{center}
\end{figure*}

The very simple methodology starts by randomly sampling a large number of parameter values from a biologically reasonable distribution (assumed uniform in this case). 
Simulations are performed with the complete set of sampled parameters, and are compared with the reference according to the distance function of Equation \ref{Eq:FitnessDenham} ordered according to their fitness.
With this methodology no intelligent search is done in the parameter space. Results obtained with this methodology are shown in Figures \ref{Fig:MC_UnifIngiVar} and \ref{Fig:MC_UnifIngiFix}. A time comparison between simulations performed with GA and MC is shown in Figure \ref{Fig:TimesSims_MC_AG}, where 
we can see that the averaged speed of GA is half the one of a simple MC brute force approach. This is because the ensemble of simulations using GA remains closer to the "real simulation" (specially for the last evolution of the algorithm) while simulations with random sampled parameters are more heterogeneous and therefore have more variability in simulation time.

 \section{More results}
 \label{Sec:moreResults}

This appendix present results using uniform initial distributions for input parameters. Each histogram shows initial distribution and  histograms built with fitted parameter values. 

\begin{figure*}
\begin{center}
\includegraphics[width=0.8\textwidth]{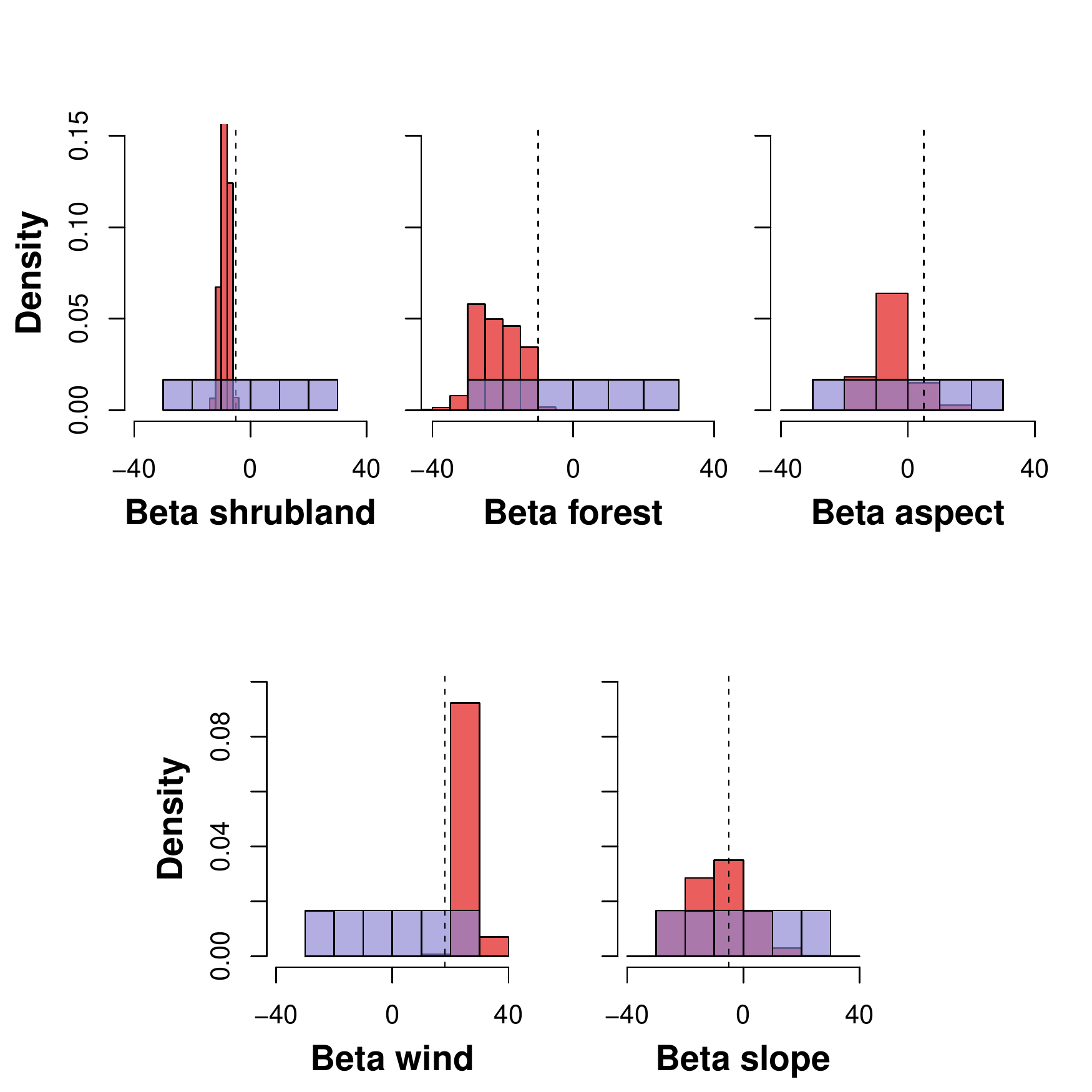}
\caption{Histograms for $\beta_i$ values after fitting the model using the genetic algorithm. Dotted line indicate the true value of the corresponding parameter that was used to generate the fire of reference. One million of simulations were performed with parameters taken from initial uniform distributions between -30 and 30 (transparent light blue). All the fires started from the same ignition point (x=400; y=250). Fitness was calculated with Equation \ref{Eq:FitnessDenham}.}
\label{Fig:IgniFixFM}
\end{center}
\end{figure*}

\begin{figure*}
\begin{center}
\includegraphics[width=0.9\textwidth]{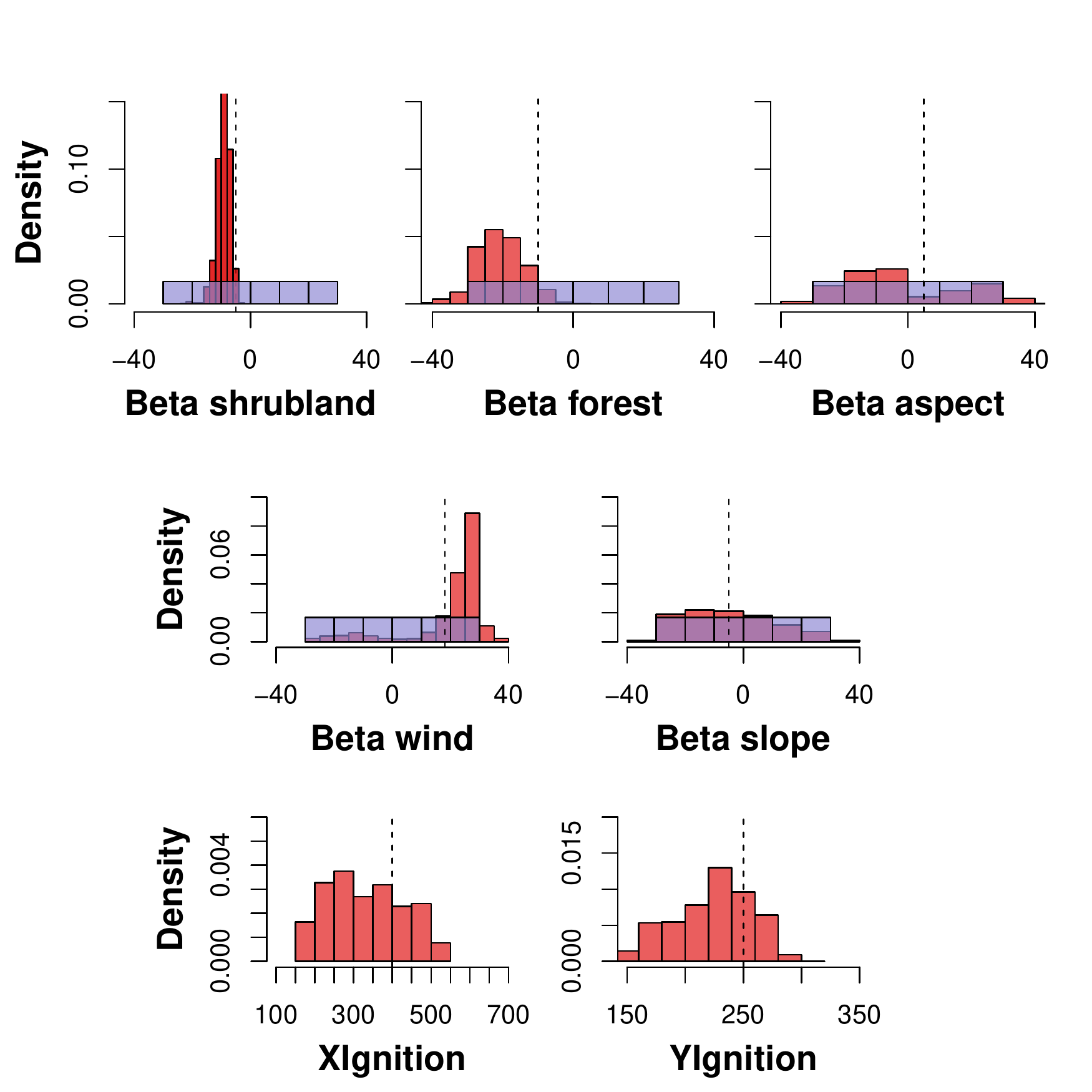}
\caption{Histograms 
for $\beta_i$ values and ignition point after fitting the model. Dotted line indicate the true value of the corresponding parameter that was used to generate the fire of reference. One million of simulations were performed with parameters taken from initial uniform distributions between -30 and 30 (transparent light blue) and fire starting from a point inside the fire scar of the fire of reference. Fitness was calculated with Equation \ref{Eq:FitnessDenham}.}
\label{Fig:IgniVarFM}
\end{center}
\end{figure*}

After performing one million simulations with parameter values sampled from a uniform distribution, we obtained histograms for each of the parameters as shown in Figures \ref{Fig:IgniVarFM} and \ref{Fig:IgniFixFM}. As can be seen in both figures, histograms of fitted parameters contain the true parameter values (Table \ref{Tab:FitPosterior}).
The best set of parameter values was defined as the one corresponding to the minimum fitness. The associated error was computed as the standard deviation of the selected ensemble of parameters. The cutoff fitness value for the selected ensemble was set as the minimum needed to obtain a stationary histogram shape for all the parameters.   
It's interesting to notice that when the ignition point is considered as an additional parameter (Figure \ref{Fig:IgniVarFM}) some parameters become less identifiable (i.e. $\beta_{slope}$ and $\beta_{aspect}$). However when the ignition point is fixed in the known values we obtain, after the fitting procedure, well defined histograms around the (known) values used for the reference fire.  

\end{document}